\documentstyle[preprint,eqsecnum,aps,prb]{revtex}
\begin{document}
\title{Ferromagnetism and metal-like transport in antiferromagnetic insulator
heterostructures.}
\author{P.\ Padhan, P.\ Murugavel and W. Prellier\thanks{%
prellier@ensicaen.fr}}
\address{Laboratoire CRISMAT, CNRS\ UMR 6508, ENSICAEN,\\
6 Bd du Mar\'{e}chal Juin, F-14050 Caen Cedex, FRANCE.}
\date{\today}
\maketitle

\begin{abstract}
Strained Pr$_{0.5}$Ca$_{0.5}$MnO$_3$/La$_{0.5}$Ca$_{0.5}$MnO$_3$/Pr$_{0.5}$Ca%
$_{0.5}$MnO$_3$ trilayers were grown on (001)-SrTiO$_3$ substrates using the
pulsed-laser deposition technique.\ The coupling at the interfaces of
several trilayers has been investigated from magnetization and electronic
transport experiments. An increase of La$_{0.5}$Ca$_{0.5}$MnO$_3$ layer
thickness induces a magnetic ordering in the strain layers and at the
interfaces leading to ferromagnetic behavior and enhanced coercivity, while
resistivity shows metal-like behaviors. These effects are not observed in
the parent compounds, which are antiferromagnetic insulators, opening a
path, to induce artificially some novel properties.
\end{abstract}

\newpage

Over the past few years, perovskite type manganites such as $R_xA_{1-x}MnO_3$
($R$=rare earth elements and $A$=alkaline earth elements) have been
extensively investigated because of their colossal magneto-resistance
properties \cite{1}, a huge decrease in the resistance with applied magnetic
field. This material has been explored in the form of thin films including
multilayer structures, made by several combinations among ferromagnetic ($FM$%
), antiferromagnetic ($AFM$), paramagnetic ($PM$) in addition to insulator
and/or metal, whose properties are different from their single layer
structures. Their physical properties have been attributed to the structural
and magnetic modifications at the interfaces of the two constituents of the
multilayers\cite{2,3,4}. For example, canted spin arrangements of
ferromagnetic layer in La$_{0.6}$Sr$_{0.4}$MnO$_3$/La$_{0.6}$Sr$_{0.4}$FeO$_3
$ and La$_{0.6}$Sr$_{0.4}$MnO$_3$/SrTiO$_3$ superlattice\cite{2}, formation
of interfacial ferromagnetism in CaMnO$_3$/CaRuO$_3$ superlattice\smallskip 
\cite{3} and disorder interfacial phase of structural and magnetic origin in
La$_{0.7}$Ca$_{0.3}$MnO$_3$/LaNiO$_3$ superlattice\cite{4} have been
observed. These examples have confirmed the importance of magnetic
interfaces which, in fact, has already been revealed in other metallic
systems\cite{5}. However, spin ordering is also observed at interfaces of
the superlattices consisting of antiferromagnetic layers of LaFeO$_3$ and
LaCrO$_3$\cite{6}. The manganites show interesting properties like the
charge/orbital ordering ($CO/OO$)\cite{7,8,9} that occurs in some half-doped
compounds. This $CO/OO$\ behavior corresponds to an ordering of charges and
orbitals in two different Mn sublattices (i.e., a long-range ordering of $%
Mn^{3+}$ and $Mn^{4+}$ ions) below the $CO/OO$ temperature, when the
materials is cooled down to low temperature. This $CO/OO$\ state is highly
an insulating state, but can be destroyed (i.e., inducing a metallic
behavior) by the application of a magnetic field\cite{8}. Similar two
prototype compounds are Pr$_{0.5}$Ca$_{0.5}$MnO$_3$ ($PCMO$) and La$_{0.5}$Ca%
$_{0.5}$MnO$_3$ ($LCMO$) exhibits insulator like behaviour with CO/OO
temperature \symbol{126}175K\cite{7} and \symbol{126}180K\cite{8},
respectively (bulk $LCMO$\ does not show $FM$ behavior\cite{9}). Here, we
report the interface effect between $PCMO$ and $LCMO$, in detail, through
transport and magnetic measurements of $PCMO$/$LCMO$/$PCMO$ trilayer
structures grown on (001)-oriented SrTiO$_3$ substrates (STO).

Thin films and heterostructures of $PCMO$ and $LCMO$ were fabricated by the
pulsed laser deposition ($PLD$) technique using a KrF laser ($\lambda =248$ $%
mm$) on STO substrates. The samples were grown at $720$ $^{\circ }C$ in an
oxygen ambient of $300$ $mtorr$\cite{10,11}. The deposition rates (typically 
$\symbol{126}$ $0.38A/pulse$) of $PCMO$ and $LCMO$ were calibrated for each
laser pulse of energy density $\symbol{126}3J/cm^2$. After the deposition
the chamber was filled to $400torr$ of oxygen at a constant rate, and then
the samples were slowly cooled down to room temperature at the rate of $%
20^{\circ }C/\min $. Trilayer structures comprising of $50$-($unit$ $cell$, $%
u.c.$) $PCMO$/$n$-($u.c.$) $LCMO$/$10$ -($u.c.$) $PCMO$, with $n$ taking
integer values from $1$ to $18$, were thusly made. To reduce the
substrate-induced strain, we have deposited a thicker bottom layer than the
top layer, while the top layer is made thinner to have the effect of
interface in the transport measurements. The structural study was done by
x-ray diffraction (XRD) using a Seifert XRD 3000P (Cu, K$\alpha 1$, $\lambda
=0.15406nm$). Special arrangement for the large intensity of x-ray beam and
large surface area of the sample was used during the $\theta $ - $2\theta $
scan. The resistivity ($\rho $) was measured using a four-probe method with 
{\it in-plane} current. The measurements are done by putting silver contact
pads, with a separation of 6 mm between the voltage electrodes of the sample
with lateral dimension close to $3\times 10$ mm$^2$. Magnetotransport and
magnetization measurements were performed with magnetic field aligned along
the [$100$] direction of the substrate. The samples were cooled to a desired
temperature from room temperature in the absence of electric and magnetic
field to perform transport measurements.

All sample shows ($00l$) fundamental Bragg's reflections of the substrate
and the constituents of the heterostructure indicate the epitaxial growth of
the trilayers. The pseudocubic lattice parameter of $STO$ ($\sim 3.905A$) is
larger than the lattice parameter of $PCMO$ ($3.802A$)\cite{7} and $LCMO$ ($%
3.83A$)\cite{8} provides {\it in-plane} tensile strain for their epitaxial
growth. The $\theta $ - $2\theta $ scan close to the fundamental ($001$)
diffraction peak of the substrate for the samples with $n=5$ and $18$ are
shown in the Fig. 1. The {\it out-of-plane} lattice parameter of various
samples with different $LCMO$\ thicknesses are shown in the inset of Fig. 1.
It also includes the bulk lattice parameters of $LCMO$ and $PCMO$. The $c$%
-axis lattice parameter of these sample increases monotonically with the
increase in $LCMO$. The $c$-axis lattice parameter of the sample with $n=18$
is lower than the lattice parameter of both $LCMO$ and $PCMO$, which
suggests the presence of substrate-induced strain state in these samples. As
a consequence, this modification in the structure of $PCMO$ and $LCMO$ with
the interfaces in the heterostructures are expected to effect the magnetic
as well as transport properties of the constituents.

Both materials, $PCMO$ and $LCMO$, are insulators with an $AFM$ behavior in
the range of our measurements\cite{10,11}. Similar insulator-like
temperature dependent resistivity $\rho $($T$) is observed in the thin films
of $LCMO$ and $PCMO$. In Fig.2, we show the zero-field cooled $\rho (T)$ in
presence of $0T$ and $7T$ in-plane magnetic field for three
heterostructures. Though the constituents of the samples are insulator the
resistivity of these samples was calculated using their actual dimensions.
As we cool the sample with 5 u.c. thick $LCMO$ layer, the zero-field
resistivity remains insulator-like down to $100$ $K$. Below $100K$, the
resistance of the sample is high and it is limited by the input impedance of
PPMS (Physical Properties Measurement System). However, for the sample with
10 u.c. thick $LCMO$, the resistivity below room temperature is
insulator-like down to $10K$. As the $LCMO$ layer thickness increases 18
u.c. and the sample is cooled from room temperature the resistivity shows
thermally activated behavior down to $150K$, shows metal-like behavior in
the temperature range of $150K$ to $30K$ and an upturn below $30K$. In
presence of $7T$ magnetic field, the $\rho $($T$) of the superlattice with 5
u.c. $LCMO$ is similar to that of the zero-field $\rho $($T$) of the sample
with 18 u.c. thick $LCMO$. Qualitatively similar in-field $\rho $($T$) with
the broader metal-like window and higher metal-insulator transition
temperature is observed for higher LCMO layer thickness. The temperature
dependent magnetoresistance ($MR$) of these heterostructures are shown in
the inset of Fig.2. The $MR$\ of both samples is decreasing as the
temperature increasing, similar to the bulk materials.

To understand this transport behavior of these heterostructures, we have
measured their magnetic properties. The temperature dependent field cooled
(0.01T) magnetization of three heterostructures are shown in the Fig.3a. The
magnetization is shown after the diamagnetic correction of $STO$. On heating
from $10K$, the heterostructure with 3 u.c. thick $LCMO$ shows a sharp
antiferromagnetic transition at $30K$ and a ferromagnetic-to-paramagnetic
transition at $250K$. As the $LCMO$ layer thickness increases the
antiferromagnetic behavior suppress and these samples show the same Curie
temperature (T$_C$). At $10K$, the magnetization of a thin film of $PCMO$ is
1.51$\times $10$^{-3}$ emu/gauss/cm$^3$ which is same order of magnetization
(9.033$\times $10$^{-3}$ emu/gauss/cm$^3$) observed for the heterostructure
with 18 u.c. thick $LCMO$.

The zero-field-cooled magnetic hysteresis loop measured at $10K$ of the
samples with $LCMO$ layer thickness above 5 u.c. shows ferromagnetic
behavior. The magnetic hysteresis loop of the sample with 18 u.c. thick $LCMO
$ layer is shown in the Fig. 3b. The magnetization of the sample increases
gradually with the increase in magnetic field and does not show a clear
saturation. We have extracted the $M_S$ (saturation magnetization) taking
into account of the weak diamagnetic response of the substrate by
extrapolating the linear part of the hysteresis loop to $\mu _0H=0$. The
extracted value of $M_S$ is $1.98$ $\mu _B/Mn$. This value of $M_S$ is small
compare the theoretical value of a ferromagnetic phase of $%
(Pr,La)_{0.5}Ca_{0.5}MnO_3$ composition ($\sim 3.5${\bf \ }$\mu _B/Mn$).
Though this sample does not exhibits a clear saturation magnetization, but
shows a significant coercive field ($H_C$) ( \symbol{126}0.07tesla).
However, the shape of the hysteresis loop does not change as we cool the
sample in presence of magnetic field. The field-cooled and zero-field-cooled
magnetic behavior of these sample does not indicate the presence of FM
cluster as expected from the $FM/AFM$ exchange bias system. The interesting
point is the origin of the $FM$ behavior, where both materials are $AFM$.
Ferromagnetic ordering of spin at the interfaces of the superlattices
consisting of two antiferromagnet $LaCrO_3$ and $LaFeO_3$ has already been
realized\cite{6}, with the ordering along the $[111]$ direction. These
samples show clear saturation in the magnetic hysteresis loop and the author
have explained $FM$ ordering due to the Goodenough-Kanamori rules. However,
to the best of our knowledge, the metal-like transport in an heterostructure
using insulators as the constituent has not been reported so far.

Several interesting behaviors have been observed in manganites by changing
average $A$-site ionic radius. This doping process provides modification in
the MnO$_6$ octahedra and formation of $FM$ metallic phase in the $AFM$
insulating matrix\cite{13,14,14a}. As seen in the inset of Fig.1, both $PCMO$
as well as $LCMO$ are in the strain state due to the lattice mismatch
between them and with the substrate\cite{11,15}. We believe that this
effective strain modify the structure as well as the spin configurations in
the $PCMO$, $LCMO$ and at their interface. However, the modification at the\ 
$PCMO/LCMO$ interface is expected to be more due to the interfacial-induced
strain and the presence of $A$-site ion La, Pr and Ca. This interfacial
stress might induce a spin re-orientation, which will modify the Jahn-Teller
distortion of the $MnO_6$ octahedra compare to the bulk materials and it may
results in spin ordering and/or spin canting at the interfaces, though
interfacial magnetic modification like spin ordering, spin frustration and
spin canting has been observed in different $FM/AFM$ systems \cite{2,3,6,16}%
. The lower value of T$_N$ in the sample with 5 u.c. thick $LCMO$, the
increase in normalized magnetization and the same value of T$_C$ with the
increase in $LCMO$ layer thickness suggest the possibility of the presence
of $FM$ domain at the interfaces due to the weakening of $CO/OO$ ordered
state, which is expected as $CE$ type ordering is most susceptible to
disorder. Since these samples show remarkable H$_C$, we believe that the
presence of $FM$ domains at the interfaces and the coupling at the $FM$
domain boundary between the $FM-AFM$ may responsible for the remarkable
value of H$_C$. The existence of $FM/AFM$ interaction should be realized in
the field-cooled hysteresis loop, but perhaps we could not able to observed
this effect due to the small interfacial volume. However, this effect was
realized once the number of interface is increasing\cite{17}.

The heterostructures with $LCMO$ layer thickness larger than 10 u.c. show a
clear metal-insulator transition in the $\rho $($T$). Also the resistivity
as well as the activation energy in the insulator-like region decreases with
the increase in $LCMO$ layer thickness. For example, the activation energy
of the heterostructure with 5 u.c. thick $LCMO$ is \symbol{126}0.12eV which
is lower than the thin film of $PCMO$ (\symbol{126}0.29eV) and $LCMO$ (%
\symbol{126}0.32eV) on $STO$. However, the activation energy of these sample
in the paramagnetic state of the $\rho $($T$) is same. The in-field
resistivity of these sample show thermally activated behavior below 30 K,
which we attribute to the grain boundary-like tunneling\cite{18}. This also
suggest the enhancement of $FM$ phase at the expense of $CO/OO$ phase in the 
$LCMO$ with the grain boundary modification\cite{8}. Since the resistance of
the sample with $n=5$ is insulator-like in the entire temperature range, we
believe that the electronic and magnetic transport in these samples are due
to spin polarized tunneling and percolative conduction as described in the
conduction process of bulk $PCMO$ and $LCMO$\cite{8,11,19}. Though there may
be some other mechanisms responsible for the metal-like conduction in the $%
\rho $($T$), we attribute it mainly to the presence of $FM$ domains near or
at the interfaces and the substrate-induced lattice distortion. The presence
of $FM$ domains may partially open double exchange conducting percolative
path, which induced the metal-like transport in these samples.

In conclusion, we fabricated epitaxial Pr$_{0.5}$Ca$_{0.5}$MnO$_3$/La$_{0.5}$%
Ca$_{0.5}$MnO$_3$/Pr$_{0.5}$Ca$_{0.5}$MnO$_3$ trilayers, where the parent
compounds are $AFM$ insulator. We observed ferromagnetic and metal-like
behavior with the increase in the La$_{0.5}$Ca$_{0.5}$MnO$_3$ layer
thickness.\ We proposed that the distribution of magnetic order and magnetic
moments near the interface are mainly responsible for the ferromagnetic
behavior. The presence of interfacial spin ordering may opens the double
exchange percolative conducting path to facilitate metal-like behavior in
these heterostructures.

We greatly acknowledge financial support of Centre Franco-Indien pour la
Promotion de la Recherche Avancee/Indo-French Centre for the Promotion of
Advance Research (CEFIPRA/IFCPAR) under Project N${{}^{\circ }}$2808-1 and
the Minist\`{e}re de la Jeunesse et de l'Education Nationale (2003/87). We
thank Dr. H.W.\ Eng for discussions.

Figure 1: Typical room temperature $\Theta -2\Theta $ x-ray diffraction
pattern around the (001) Bragg's peak of (50 u.c.) $PCMO/$(n u.c.) $LCMO/$%
(10 u.c.) $PCMO$ trilayer with (a) $n=5$ and (b) $n=18$ grown on
(001)-oriented STO. The inset shows the {\it out-of-plane} lattice parameter
of various trilayer structures with different $LCMO$ layer thicknesses. The
bulk lattice parameters of $PCMO$ and $LCMO$\ are also indicated.

Figure 2: Zero-field-cooled $\rho (T)$ in the presence of $0$ tesla (open
symbols) and $7$ $tesla$\ (full symbols) magnetic field for different
trilayers. The inset depicts the $MR$ ($MR=\frac{\rho (0)-\rho (7T)}{\rho
(7T)}$) for the heterostructures with $n$ $=$ $5$ and $18$. Note that under
magnetic field, all curves show an insulator-to-metal transition.

Figure 3(a): Field-cooled temperature dependent magnetization of different
heterostructures at $10K$ at 0.01 tesla magnetic field.\ (b)
Zero-field-cooled magnetic field dependent magnetization at $10K$ of the
heterostructure with $18$ $u.c.$ thick $LCMO$.

\end{document}